\documentstyle[twocolumn,prc,aps]{revtex}
\input epsf

\begin{document}

\twocolumn [ \hsize\textwidth\columnwidth\hsize\csname @twocolumnfalse\endcsname
  
  \title {Periodic $\theta$ Vacuum Model and the QCD Instanton Vacuum }

  \author {
              	Prashanth Jaikumar and Ismail Zahed
  }  
\address {
     Department of Physics and Astronomy, State University of New York, 
     Stony Brook, NY 11794-3800 
  }
  \date{\today}
  \maketitle    

\begin{abstract}
We formulate a matrix model for the $\theta$ vacuum that
embodies the n-point moments of the even and odd topological
charge densities. The partition function is $2\pi$ periodic
only after resumming over all winding modes. We apply the
same analysis to the quenched instanton vacuum at finite 
$\theta$ and derive saddle-point  results for the vacuum energy 
density that are manifestly $2\pi$ periodic whatever $\theta$.
We briefly discuss the consequences of unquenching the fermions
on our results.
\vskip 0.2cm
PACS number(s): 12.38.Aw, 11.15.Tk, 11.30.-j.
\end{abstract}
\vspace{0.1in}
]

{\bf 1.\,\,}
The non-abelian nature of QCD allows for the existence of classical,
finite action, large gauge configurations that interpolate continuously 
between topologically inequivalent vacua. The canonically quantized 
vacuum state is quasi-periodic under large gauge transformations, with
$\theta$ playing the role of the quasi-momentum. States with different
$\theta$ correspond to different worlds. The dual to $\theta$ 
is the topological charge and large gauge configurations of unit 
topological charge are instantons~\cite{POLYAKOV,THOOFT}.

In the past years, the instanton approach to the QCD vacuum and its 
low-lying excitations has received considerable vindication both 
theoretically and numerically~\cite{REVIEW}. In the instanton vacuum, 
$\theta$ plays the role of an imaginary chemical potential, weighting 
incoherently the gauge configurations of net topological charge. As a 
result, most importance sampling arguments fail at finite $\theta$.
For small $\theta$, theoretical arguments based on the instanton
vacuum yield a nontrivial $\theta$ dependence of the vacuum partition
function~\cite{DIA}. However, for large $\theta$ the results 
lack explicit $2\pi$ periodicity, a point first addressed in~\cite{WITTEN}.

In this study, we show how to construct a partition function for the
quenched instanton vacuum that is manifestly $2\pi$ 
periodic whatever $\theta$, following recent and general suggestions
in~\cite{hz}. In section 2, we illustrate this approach by showing
how the $2\pi$ periodicity is enforced in a matrix model of the QCD partition function, making the arguments
of~\cite{hz} transparent. In section 3, we extend these arguments to
the instanton vacuum model. In the quenched case, we derive explicit 
expressions for the instanton vacuum energy whatever $\theta$. For
small $\theta$ our results reproduce those in~\cite{Weiss}. In
section 4, we extend our analysis to the unquenched case both in
the matrix model and instantons. Our conclusions are in section 5.

\vskip 1.25cm
{\bf 2.\,\,}
As finite $\theta$ angle simulations for QCD are not feasible on the lattice, a study of the $\theta$-dependence is more accessible through a model such as the matrix model that can mock up the essentials of the $\theta$ vacuum. The matrix model for the $\theta$ vacuum was studied in ~\cite{Z} with gaussian weights for the even and odd topological charge densities. Results from that model calculation directly address the levelling of the free energy as a function of $\theta$, finite-size effects and effects of precision on the numerical analysis. However, we can inject more information into the model by using the low energy theorems to ensemble average with non-gaussian weights, as follows.
In QCD, the connected 
2-point correlation function for the scalar glueball obeys
the low-energy theorem~\cite{NSVZ}
\begin{equation}
\small{i\int\, d^4x\langle T^* { \frac{\beta}{4\alpha}GG(x)  
\frac{\beta}{4\alpha}GG(0) }\rangle = 
-4\langle\frac{\beta}{4\alpha}GG (0)\rangle} \label{W1}\,\,,
\end{equation}
where $\alpha$ is the strong coupling constant, and $\beta$ is the
1-loop beta function. The 2-point correlation function for the 
pseudoscalar and scalar-pseudoscalar mixing obey similar low-energy
theorems~\cite{NSVZ,hz2}
\begin{equation}\small{i\int\, d^4x\,\langle 
T^* { \frac{\alpha}{8\pi}G\tilde{G}(x)\frac{\alpha}{8\pi}G\tilde{G}(0)
}\rangle 
={\xi^2}\langle\frac{\beta }{4\alpha}\,GG(0)\rangle} \label{W2}
\end{equation}
\begin{equation}
\small{i\int d^4x\,\langle
T^*{ \frac{\alpha}{8\pi}GG(x)\frac{\alpha}{8\pi}G\tilde{G}(0) }\rangle 
= 2\xi\langle\frac{\alpha}{8\pi}G\tilde{G}(0)\rangle} \label{W3} \,\,, 
\end{equation}
where ${2\xi={q}/{p}={4}/{b}}$ for the quenched case with ${b=-2\pi\beta/\alpha^{2}={11N_{c}}/{3}}$. We note that in non-SUSY Yang-Mills theory, the precise values of $p,q$ beyond one-loop are still debated~\cite{Hal}, therefore we refer to them generally as $p$ and $q$ in what follows.
Higher correlation functions using the scalar and pseudoscalar sources
can be defined using similar arguments~\cite{hz2}. 
\vskip 0.1cm
The matrix ensemble used to carry out the averaging has sizes $n_{+}\times n_{-}$, which relate directly to the low energy theorems, thanks to the identification
\begin{equation}
n_\pm \approx \frac \alpha{16\pi}\,\int \,d^4x\, \left( GG(x) \pm iG\tilde G
(x)\right)\,\,,
\label{chiral}
\end{equation}
which is a standard identification in matrix models of the instanton vacuum where $n_{\pm}$ counts the number of right-handed and left-handed zero modes (restricted to zero dimension) respectively.
Then the low-energy theorems ~Eqs.(\ref{W1}),(\ref{W2}),(\ref{W3}) translate to conditions on the n-moments
\begin{equation}
\small{\langle  (n_{\pm})^{n+1}\rangle =(\frac{q}{p})^{n}\langle n_{\pm} \rangle}\,\,.
\label{moment}
\end{equation}
  Given the connected $n$-moments (\ref{moment}), we can construct the associated generating function as
\begin{equation}
{W(J_{-},J_{+}) = \frac{p}{q}\langle n_{-}\rangle{\rm 
exp}(4J_{-})+\frac{p}{q}\langle n_{+}\rangle{\rm exp}(4J_{+})}\,\,,
\label{generating}
\end{equation}
where ${J_{-},J_{+}}$ are pertinent sources.
At finite vacuum angle $\theta$, the generating function
(\ref{generating}) changes. Since the low-energy theorems
(\ref{moment}) also hold in the $\theta$ vacuum, the
$\theta$ dependence amounts to a change in the sources $J_\pm$
\cite{hz}. This can be seen as follows. The $\theta$ parameter enters as the coefficient of the topological charge
\begin{equation}
{\cal L}_{\theta}=frac{\theta\\alpha}{16\pi}\int d^4x\,G\tilde{G}(x)\quad.
\end{equation}
In terms of the matrix model, it is a coefficient of the topological susceptibility $\chi\sim (n_{+}-n_{-})$. When exponentiated to obtain the generating function, it thus serves to shift the sources for both $n_+$ and $n_-$, i.e. $J_+$ and $J_-$ as $J_{\pm}\rightarrow J_{\pm}\pm i\xi\theta/2$. Since $\xi\sim 1/N_c$, the dependence of the partition function on $\theta$ appears as {\mbox{cos}}\,($\theta/N_c$) (see below), which is clearly not periodic in $\theta$. 
\begin{equation}
Z(\theta)=e^{W_{\theta}[J_+,J_-]}={\mbox{exp}}\biggl(\frac{p}{q}\langle n_++n_-\rangle{\mbox {cos}}\,(\frac{q}{p}\theta)\biggr)\quad,
\end{equation}
where we have used $<n_+-n_->=0$. This problem precisely was addressed in~\cite{hz}. A solution of the $U(1)$ problem in QCD requires that the $\theta$ dependence does appear in this form~\cite{WITTEN}. However, physical quantities should be $2\pi$ periodic  in $\theta$ since it is a vacuum angle. This problem was identified by the authors of~\cite{hz} by noting that the infinite volume limit taken in calculating physical quantities like the free energy does not commute with the $2\pi$ shift in $\theta$, and that it can be cured by imposing the constraint that the topological charge can take only integer values. This explicit global equation of constraint does not affect the physics extracted from the partition function defined in such a way. Therefore, 
to enforce manifest $2\pi$ periodicity in $\theta$ in our model,
we follow the arguments in~\cite{hz}, and use the identity
\begin{equation}
\sum_{{\bf m}=-\infty}^{+\infty}\,\,\delta ( \frac qp (n_+-n_-)  -{\bf m} ) =
\sum_{k=-\infty}^{+\infty} \,e^{2\pi\,ik\ \frac qp (n_+-n_-)} \quad.
\label{identity}
\end{equation}
Note that this condition involves the integers $p$ and $q$ defined previously. All we are doing is introducing a divergent normalization $\sum 1$ to the definition of the partition function. However, this is not a trivial step since $n_+$ and $n_-$ are restricted to certain values satisfying the above condition. With this explicit summation, the partition function becomes
\begin{equation}
\small{Z(\theta) =\sum_{k=-\infty}^{\infty}\exp\,\biggl(\frac{p}{q}\langle n_{+}+n_{-} 
\rangle\cos(\frac{-q}{p}\theta+2\pi\frac{kq}{p})\biggr)}\,\,,
\end{equation}
We may rewrite this as
\begin{equation}
\small{Z(\theta) =\sum_{l=0}^{p-1}\exp\,\biggl(\frac{p}{q}\langle n_{+}+n_{-} 
\rangle\cos(\frac{-q}{p}\theta+2\pi\frac{l}{p})\biggr)}\,\,,
\label{partition}
\end{equation}
using the $2\pi$ periodicity of the cosine. 
The vacuum energy at finite $\theta$ follows from ${E}=W/V=-{\rm ln}Z/V$. Specifically,
\begin{equation}
{E} (\theta)  = \lim_{V\rightarrow\infty} 
\frac{-1}{V}\ln\,\left(\sum_{l=0}^{p-1}\mbox{exp}(VE (0)\,\cos(\frac{-q}{p}\theta+
2\pi\frac{l}{p}))\right)
\end{equation}
which is exactly the closed form version of the vacuum energy discussed
in~\cite{hz} using different methods. Here, $E(0)$ is the vacuum energy at $\theta=0$. In fact, this is the functional form that must be
obtained if the holomorphic combinations ${{G}G\pm\,i\, G\tilde{G}}$
are selected to obtain the moments. In other words, the cosine is generic.
The non-commutativity of the thermodynamical limit and the shift ${\theta 
\rightarrow \theta + 2\pi}$ is a consequence of the multi-valuedness over 
${l}$, as pointed out first in ~\cite{hz}. This means that periodicity in ${\theta}$ can be reconciled with the thermodynamical limit only if we select the right values of ${l}$. In simple terms, we encounter different ``Reimann sheets'' in making the 
${2\pi}$ shift so that the entire branch structure must be
retained to account for the periodicity in ${\theta}$. Armed with a partition function for the matrix model that has the correct periodicity in $\theta$, we can comment on the features that would arise in a numerical simulation of such a model. We stress that a matrix model is not intended to represent the dynamics of the QCD vacuum. Rather, we have written down a partition function within a toy model that is amenable to numerical simulations which would yield the qualitative dependence of the free energy on $\theta$.

In anticipation of possible numerical simulations of the partition
function at finite $\theta$, it is useful to analyze the partial
partition functions for configurations with $N={\rm max}\, n_\pm$.
Specifically,
\begin{eqnarray}
Z_{N,l}(\theta) =\exp(VE (0)\,\cos(-\frac{q}{p}\theta+\frac{2\pi 
l}{p})) \nonumber  \\ 
\times\frac{\Gamma(N+1,\mu)}{N!}\frac{\Gamma(N+1,\nu)}{N!}\,\,,
\label{partial}
\end{eqnarray}
where
\begin{equation}
\small{\mu = \langle 
n_{+}\rangle\frac{p}{q}\exp(-i\frac{q}{p}\theta+\frac{2\pi l}{p})}\,\,
\end{equation}
and
\begin{equation}
\small{\nu = \langle 
n_{-}\rangle\frac{p}{q}\exp(i\frac{q}{p}\theta+\frac{2\pi l}{p})}\,\,.
\end{equation}
Here, ${\Gamma}$ is the incomplete gamma function. Note that for
$N\rightarrow \infty$ we recover the partition function (\ref{partition}).
Recently, it was argued by a number of authors~\cite{Z,SCH,PS} that 
a lattice assessment of the vacuum energy at finite $\theta$ in
CP$^n$ models and Yang-Mills theory show an unexpected levelling 
at large $\theta$ with consequences on the strong CP problem. However,
this observation may be a numerical artifact~\cite{Z,PS}. In our case, 
this issue can be readily investigated  by noting that (\ref{partial}) 
approximates the full partition function at finite but large $N$.
Numerically, corrections in $1/N$ start to be
appreciable when
\begin{equation}
\small{{\rm ln}\,({q}/{pn_{*}})+n_{*}\frac{p}{q}\cos({q}\theta/p ) \leq 1}\,\,,
\end{equation}
where ${n_{*}}=N/V$ is the maximum density of winding modes. For
simplicity, we have specialized to the $l=0$ branch for which the
ground state occurs for 
${0 \leq \theta \leq \frac{\pi}{q}}$. For fixed ${p}$ and ${q}$, the 
deviations between the results for $N=\infty$ and $N$ finite
decrease with increasing ${n_{*}}$. For fixed ${q}$ and ${n_*}$, 
the deviations decrease with increasing ${p}$ (recall that ${p \sim
N_{c}}$). The changes in the $p, q, n_*$ parameters
in the finite sum (\ref{partial}) can significantly alter the 
$\theta$ dependence of the vacuum energy~\cite{Z,PS}.
These observations are useful for the lattice calculations 
carried in~\cite{SCH,PS}. We now move to applying this periodicity procedure for $\theta$ to the interacting instanton ensemble.
\vskip 1.25cm
{\bf 3.\,\,}
The prescription for ensuring 2${\pi}$ periodicity for the model vacuum
partition function discussed above, can be readily extended to a
weakly interacting ensemble of instantons and anti-instantons~\cite{REVIEW,DIA}.
For that, we consider the vacuum partition function at finite $\theta$
(vacuum angle) and $\mu$ (chemical potential for winding modes) from~\cite{Weiss}
\begin{equation}
Z(\mu, \theta) = \sum_{N_{\pm}}
\mbox{e}^{\mu (N_{+}+N_{-})} 
\mbox{e}^{i\,\theta (N_{+}-N_{-})} Z_{N_{\pm}}\,\,.
\label{INS1}
\end{equation}
The partial sums are given by
\begin{equation}
 Z_{N_{\pm}} = \frac{(V\zeta_+)^{N_+}}{N_{+}!} 
\frac{(V\zeta_-)^{N_-}}{N_{-}!} \prod_{\langle ij\rangle}^{N_{\pm}}
 \int d{ R}_i\,d{ R}_j \,\,e^{-{\bf S} (R_i, R_j)}\,,
\label{INS2}
\end{equation} 
where $\zeta_\pm$ are the pertinent fugacities~\cite{DIA}.
The instantons are characterized by a set of collective
coordinates ($R$), such as their position, color orientation and size
($\rho$). 
For simplicity, we choose the relative two-body action
in the schematic form~\cite{DIA,Weiss}
\begin{eqnarray}
{\bf S} ( R_1,R_2) 
 \approx \frac{\lambda}{V}\,\gamma_{12}\,( {\rho_{1}}\, {\rho_{2}})^{2} \,\,,
\nonumber
\end{eqnarray}
as suggested by variational calculations. Here ${\gamma}_{12}=\gamma_s$ 
for like particles, and $\gamma_{12}=\gamma_a$ for unlike ones.
${\lambda}$ is the inverse charge at the typical
relative separation and $V$ is the 4-volume. 
The ${\theta}$ dependence of the vacuum energy is easy to write down
for special values of ${r={\gamma_{a}}/{\gamma_{s}}}$, as pointed
out in \cite{Weiss}. For ${r = 1}$, (which is the sum ansatz 
of~\cite{DIA})
\begin{equation}
f(\theta) = {[\mbox{cos}(\theta)]}^{\frac{4}{b}}\,\,,
\end{equation}
where ${f(\theta)}$ expresses the ${\theta}$-dependence of the 
partition function written in a generic form as
\begin{equation}
Z(\mu,\theta) = {\rm{exp}}(\frac{b}{4}\langle N\rangle_{0} 
f(\theta){\rm{exp}}(\frac{4\mu}{b}))
\end{equation}
and ${\langle N\rangle_{0}}$ is the mean instanton($I$) and
anti-instanton($\bar{I}$) number at zero vacuum angle~\cite{DIA}.
For ${r = 0}$,
\begin{equation}
f(\theta) = \mbox{cos}(\frac{4\theta}{b})\,\,,
\end{equation}
and finally, for ${ r \rightarrow -1}$,
\begin{equation}
f(\theta) = {[\mbox{cos}\left(\frac{2\theta}{b-2}\right)] }^{\frac{2(b-2)}{b}}\,\,.
\end{equation}
The key point we notice is that for $r=0$ and $r\rightarrow -1$, the free energy is not $2\pi$ periodic in $\theta$. The problem and it's solution are clear once we see that the free energy is extracted following a saddle-point analysis
discussed in~\cite{Weiss}. It involves the relative
ratio
\begin{equation}
\Phi = \frac{{\rm ln}({N_{+}}/{N_{-}})}{2i}\,\,,\nonumber \\
\end{equation}
which satisfies the saddle-point equation
\begin{equation}
 \Phi + \frac{b-4}{4}\mbox{arctan}\frac{(1-{r}^{2})\mbox{sin}\,\Phi}{\mbox{cos}\, \Phi + 
r\sqrt{1-{r}^{2}{\mbox{sin} }^{2}\,\Phi}} = \theta\,\,.
\end{equation} 
A shift of $2\pi$ in $\theta$ can be compensated by a similar shift in $\Phi$, so $2\pi$ periodicity seems implicit here. However, the free energy involves taking a thermodynamic limit which does not commute with the $2\pi$ shift as explained before. This is why the free energy expressions are not $2\pi$ periodic in $\theta$ (except for $r=1$ which is a special case discussed later). If the free energy, which is a physical quantity, has to be $2\pi$ periodic in $\theta$, the explicit but unconventional sum over winding modes
has to be enforced as before, and the modified partition function is
\begin{equation}
Z(\mu, \theta) = \sum_{k=-\infty}^{+\infty} \sum_{N_{\pm}}
\mbox{e}^{\mu (N_{+}+N_{-})} 
\mbox{e}^{i\,(\theta +2 \pi k) (N_{+}-N_{-})} Z_{N_{\pm}}\,\,.\label{New}
\end{equation}
The fixed ${N_{\pm}}$ partition function can
be varied to determine the best size
distribution. Evaluation of the partition function requires an
expression for the average size as well, and these can be determined
from consistency equations. The full partition function can then be
calculated in a saddle-point analysis. We refer the reader to~\cite{DIA}
for a further and explicit discussion and state here that the new saddle point equation becomes
\begin{equation}
 \Phi + \frac{b-4}{4}\mbox{arctan}\frac{(1-{r}^{2})\mbox{sin}\,\Phi}{\mbox{cos}\, \Phi + 
r\sqrt{1-{r}^{2}{\mbox{sin} }^{2}\,\Phi}} = \theta + 2\pi k\,\,.\label{N0}
\end{equation} 
The infinite sum over ${k}$ looks superfluous but it is not. It implies an infinite number of saddle points, but
as before, this may be rewritten as a principal solution in a branch ${l}$,
plus an infinite number of copies, that can be dropped because they all
yield the same minimum. This also explains why the
summation over ${l}$ is finite ( 0 to $p-1$ ). The value of ${l}$ to choose
depends on the range of ${\theta}$ and the integers ${q}$ and
${p}$. These observations are novel and absent in~\cite{Weiss}. If we ask the question of $2\pi$ periodicity of $\theta$ {\it after} performing the thermodynamic limit, it is essential that this limit is taken properly, that is, with all branches described by $l$ taken into account. When the thermodynamic limit is taken properly, the character of the vacuum energy at finite $\theta$ is changed and is explicitly periodic as we now
show. For ${r = 0}$,
\begin{equation}
f(\theta) = \mbox{cos}(\frac{4\theta}{b}+\frac{2 \pi l}{p})\,\,,\label{N1}
\end{equation}
where ${l}$ may be any integer between 0 and $p-1$, depending on 
which range ${\theta}$ lies in. To see the $2\pi$ periodicity, one has to retain all appropriate branches of l corresponding to the different ranges of $\theta$ as it moves through $2\pi$. Then we observe that the soutions are merely relabelled but the entire set in $l$ remains the same. This is the subtle way in which periodicity in $\theta$ is enforced.  
Finally, for ${ r \rightarrow -1}$, we get
\begin{equation}
f(\theta) = {\biggl[\mbox{cos}\left(\frac{2\theta}{b-2}+
\frac{2\pi l}{p}\right)\biggr]}^{\frac{2(b-2)}{b}}\,\,. \label{N2}
\end{equation}
Here too, ${l}$ lies between 0 and $p-1$, but the ranges for these 
choices is different because of the different cosine dependence. 
This is a limiting case since a stable ensemble does not exist for ${r
\leq -1}$. ~Eqns(\ref{N0}),(\ref{N1}),(\ref{N2}) are new and flow from the modified partition function~Eq.(\ref{New}). The case $r=0$ is special since
in this case, the expression is periodic without the ${2\pi k}$ which is
indeed superfluous. This seems to indicate that the free energy depends only on the total number of pseudoparticles in the system, and is insensitive to the difference in the number of instantons and anti-instantons. This would be natural since the sum ansatz sets the $II$ interaction equal to the $I\bar{I}$ interaction. In conclusion, we see that the results of~\cite{Weiss} admit a generalization in our
approach which makes them manifestly periodic in the same way as 
in the model discussed above.
Lattice cooling measurements with rich instanton content  
should be useful in assessing the ${\theta}$ dependence of the vacuum
energy along the present lines, including the value of $r$.

\vskip 1.25cm
{\bf 4.\,\,}
In so far, both the model and instanton analysis of the energy of the
$\theta$ vacuum was carried without fermions (quenched). The addition
of fermions modify substantially our results. Given the similarities
between the model and the instanton analysis, we will show in this
section how our model calculation is affected by the introduction of fermions.
This is of course a simplified version of a full fermion analysis in an instanton ensemble, but should be useful for an unquenched lattice simulation nevertheless.
We will derive analytic results for the vacuum energy as a function of $\theta$ for small fermion masses in the cases $N_f=1,2$. From this, the topological susceptibility  can be determined and our model can be confronted with experimental cooling simulations on a lattice.
The addition of fermions (${N_{f}}$ flavors) modifies the model
partition function as follows~\cite{Z,Now}
\begin{equation}
\small{Z(\theta,N_{f})=\langle\prod_{j=1}^{N_{f}}\det(\begin{array}{cr}
im_{j}e^{i\frac{\theta}{N_{f}}} & W\\
W^{\dag}                           & im_{j}e^{-i\frac{\theta}{N_{f}}}
\end{array})\bf{\rangle}}
\end{equation}
where W is a complex hermitian asymmetric matrix of dimension ${k_{+}\times 
k_{-}}$ (not the same as ${n_{+},n_{-}}$). The reason that $k_{\pm}$ is not the same as $n_{\pm}$ has to do with the chiral anomaly, as we will see. For simplicity, the measure on
$W$ is dictated by the maximum entropy principle, hence gaussian. 
The quark mass matrix is introduced in a form consistent with the 
chiral anomaly
\begin{equation}
\theta\longrightarrow\theta+\arg  \det M\,\,.
\end{equation}
To perform the disorder average successfully, we must make a modification in 
the process of bosonization, that will determine the weight to be used. In the 
presence of fermions, the explicit $2\pi$ periodicity is enforced by
requiring that the matrix asymmetry satisfies
\begin{equation}
k_{+}-k_{-}={\bf m}\,\,,
\end{equation}
with ${\bf m}$ fixed by the additional identity (\ref{identity}).
This constraint implies that the
number of zero modes is not the same as the number of winding
modes for the unquenched case, and is essential to preserve the 
form of the chiral anomaly. Now, bosonization may be performed in a 
manner consistent with the anomaly, and the unquenched partition 
function reads
\begin{eqnarray}
Z(\theta,N_{f})= \sum_{\bf{k}}\int{dPdP^{\dag}}{\mbox{e}^{-\frac{\bf{k}}{2}tr 
P^{\dag}P+{\bf{k}}\hspace{0.02in}{\rm Tr}\hspace{0.02in}{\rm log}|m+P|}} \nonumber\\
\times \sum_{l=0}^{p-1}{\rm
 e}^{ VE (0) \cos(\frac{-q}{p}(\theta+\phi)+2\pi\frac{l}{p})} 
\end{eqnarray}
where ${{\bf{k}}=k_{+}+k_{-}}$ and ${\phi={\rm 
arg\hspace{0.02in}det\hspace{0.02in}}M}$.
Once again, the periodicity in ${\theta}$ is realized over the set of all 
branches of the function, so we pick a particular value of ${l}$ for each 
range of ${\theta + \phi}$. For example,
when ${0\leq(\theta+\phi)\leq {\pi}/{q}}$, the ${l=0}$ 
branch dominates in the thermodynamic limit. A saddle-point
approximation to 
the partition function may be carried out with 
$P={\rm diag}\,\,p_{j}{\rm  e}^{-i(\phi_{j}-{\theta}/{N_{f}})}$, (where $j$ labels flavor) to 
yield the unquenched vacuum energy.

Considering the range of ${\theta}$ for which ${l=0}$ 
dominates and for small masses ${m_{j}\ll p_{j}}$, we
obtain for the unquenched vacuum energy
\begin{eqnarray}
{\cal E} (\theta ) =\frac{k_{*}}{2}\sum_{j=1}^{N_{f}}\,\,\biggl([p_{j}^2-log\hspace{0.1in}p_{j}^2+2\frac{m_{j
}}{p_{j}}\cos\,\phi_{j}] \nonumber\\
+E (0) \,\cos(-\frac{q}{p}\theta+\frac{q}{p}{\phi_{j}}+
\frac{m_{j}}{p_{j}}{\rm sin}\,\phi_{j})\biggr) 
\end{eqnarray} to ${{\cal{O}}(m^2)}$,
where we have taken ${\bf{k}}$ to be peaked around ${{k_{*}}}$.
Thus, we have written down a distribution in ${k_{+}+k_{-}}$ apart 
from the natural one in ${k_{+}-k_{-}}$.
 The saddle points in the ${p}$'s decouple to give
\begin{equation}
p_{j}=1-(\frac{m_{j}}{2})\cos\,\theta+{\cal{O}}(m^2) 
\end{equation}
and to the same order, the saddle points in the ${\phi}$'s is
\begin{equation}
\theta=\sum_{j=1}^{N_{f}}\phi_{j}+{\cal{O}}(m)\,\,.
\end{equation}
It is easy to show that
\begin{equation}
m_{1}\sin\,\phi_{1}=m_{2 }\sin\,\phi_{2}=...m_{N_{f}}\sin\,\phi_{N_{f}}\,\,.
\end{equation}
These approximate expressions were obtained in the large ${N_{c}}$ limit as 
well \cite{WITTEN}. Note that the topological susceptibility
$\chi$ for equal quark masses follows from
\begin{equation}
{\cal E}_{min} (\theta) =-E (0) -k_{*}\,mN_{f}\,\cos({\theta}/{N_{f}})+
{\cal O} (m^2)
\end{equation}
in the form
\begin{equation}
\chi= \frac{d^2}{d\theta^2}{\cal E}_{\rm min} (\theta)|_{\theta=0}=k_{*}\frac{m}{N_{f}}+{\cal{O}}(m^2)\,\,,
\end{equation}
a well-known result~\cite{WITTEN}. 

In general, for the lth branch
\begin{equation}
{(2l-1)\frac{\pi}{q}\leq(\theta-\sum_{j=1}^{N_{f}}\phi_{j})\leq 
(2l+1)\frac{\pi}{q}}\,\,,
\nonumber
\end{equation}
and unequal masses, we have
\begin{eqnarray}
{\cal E} (\theta) =-E(0)
{\rm cos}\biggl(-\frac{q}{p}\theta+\frac{q}{p}\sum_{j=1}^{N_{f}}\phi_{j
}+2\pi\frac{l}{p}\biggr) \nonumber\\
-k_{*}\sum_{j=1}^{N_{f}}m_{j}\cos\phi_{j}\,\,.
\end{eqnarray}
For ${N_{f}=1}$, we have
\begin{equation}
\phi=\theta+m\biggl(\frac{k_{*}}{E(0)}(\frac{p}{q})^2-1\biggr)\sin\,\theta+O(m^2)\,\,,
\end{equation}
for which the subtracted vacuum energy reads
\begin{equation}
\Delta {\cal E} (\theta) =k_{*}\,m\,(1-\cos\hspace{0.01in}\theta)\,\,.
\end{equation}
Although the branch-structure affects the position of the saddle point,
it does not produce a physical change in the vacuum energy.
At small ${m}$, for any ${\theta}$, there is no cusp. There would 
be a cusp at ${\theta = \pi}$ if ${m}$ were large~\cite{Z} but the saddle point analysis is valid only for small ${m}$.
For ${N_{f}=}$ 2, we have
\begin{equation}
\sin\phi_{1,2}={\pm}\frac{m_{2,1}\sin\theta}{\sqrt{m_{1}^2+m_{2}^2+2m_{1}m_{2}\cos\theta}}\,\,,
\end{equation}
with the subtracted vacuum energy
\begin{equation}
\Delta {\cal E}(\theta)=k_{*}[\hspace{0.02in}\mid 
m_{1}+m_{2}\mid-\sqrt{m_{1}^2+m_{2}^2+2m_{1}m_{2} \cos\,\theta}\hspace{0.04in}]\,\,.
\end{equation}
For ${m_{1}=m_{2}}$, the free energy has a cusp at ${\theta=\pi}$, which 
indicates a spontaneous breaking of strong CP.

Most of the above results apply to an ensemble of weakly interacting
instantons and anti-instantons as discussed above, since the
pair-interaction is ``screened'' by the quarks through the
fermion determinant. The multivalued structure of the 
unquenched partition function should be similar to what is shown in this section.

\vskip 1.25cm
{\bf 5.\,\,}
In conclusion, following on recent suggestions~\cite{hz}, we have shown how
a multivalued approach to the $\theta$ vacuum can be displayed
in a simple model, deriving closed form results for the vacuum
energy that are manifestly periodic whatever $\theta$. This model 
is particularly useful for testing the pertinence of numerical
simulations in a field theory (CP$^n$ and Yang-Mills) at finite
$\theta$. We have used the insights generated by the analytically
tractable model to derive the vacuum energy at finite $\theta$ 
of a quenched and dilute instanton anti-instanton ensemble, that 
is manifestly $2\pi$ periodic, generalizing earlier
results~\cite{Weiss}. While direct lattice calculations at finite $\theta$ are not feasible due to the complex integration measure, quenched and cooled simulations can test the energy dependence on $\theta$ via a measurement of the topological charge. We have shown how
the fermions affect the model calculation, and suggest that 
the same generic results apply to the unquenched instanton
calculation. Our analysis can be improved, by considering 
$\xi=q/p$ beyond the one-loop result, still a point of
discussion~\cite{Hal}.

\section*{Acknowledgements}
This work was supported by the US-DOE grant DE-FG-88ER40388.
\begin{flushleft} 

\end{flushleft}
\end{document}